\documentclass[useAMS,usenatbib,usegraphicx]{mn2e}

\title[Stripping of nitrogen-rich AGB ejecta ]{Stripping of nitrogen-rich AGB ejecta from interacting dwarf irregular galaxies}
\author[T. Tsujimoto and K. Bekki]{Takuji Tsujimoto$^{1,2}$\thanks{E-mail:
taku.tsujimoto@nao.ac.jp} and Kenji Bekki$^{3}$\\
$^{1}$National Astronomical Observatory of Japan, Mitaka, Tokyo 181-8588, Japan\\
$^{2}$Department of Astronomical Science, The Graduate University for Advanced Studies, Mitaka, Tokyo 181-8588, Japan\\
$^{3}$ICRAR, M468, The University of Western Australia, 35 Stirling Highway, Crawley Western Australia 6009, Australia\\}

\def\ltsima{$\; \buildrel < \over \sim\;$}
\def\ltsim{\lower.5ex\hbox{\ltsima}}
\def\gtsima{$\; \buildrel > \over\sim \;$}
\def\gtsim{\lower.5ex\hbox{\gtsima}}
\def\ms{$M_{\odot}$ }
\def\msp{$M_{\odot}$}

\begin{document}

\date{Accepted 2013 August 29. Received 2013 August 7; in original form 2013 June 7}

\pagerange{\pageref{firstpage}--\pageref{lastpage}} \pubyear{2013}

\maketitle

\label{firstpage}

\begin{abstract}
Dwarf irregular galaxies (dIrrs) including the Magellanic Clouds in the local Universe, in many cases, exhibit an unusually low N/O abundance ratio ($\log$ N/O$\sim-1.5$) in H II regions as compared with the solar value ($\sim-0.9$). This ratio is broadly equivalent to the average level of extremely metal-poor stars in the Galactic halo, suggesting that N released from asymptotic giant branch (AGB) stars is missing in the present-day interstellar matter of these dIrrs. We find evidence for past tidal interactions in the properties of individual dIrrs exhibiting low N/O ratios, while a clear signature of interactions is unseen for dIrrs with high N/O ratios. Accordingly, we propose that the ejecta of massive AGB stars that correspond to a major production site of N can be stripped from dIrrs that have undergone a strong interaction with a luminous galaxy. The physical process of its stripping is made up of two stages: (i) the ejecta of massive AGB stars in a dIrr are first merged with those of the bursting prompt SNe Ia and pushed up together to the galaxy halo, and (ii) subsequently through tidal interactions with a luminous galaxy, these ejecta are stripped from a dwarf galaxy's potential well. Our new chemical evolution models with stripping of AGB ejecta succeed in reproducing the observed low N/O ratio. Furthermore, we perform N-body + hydrodynamical simulations to trace the fate of AGB ejecta inside a dIrr orbiting the Milky Way, and confirm that a tidal interaction is responsible for the efficient stripping of AGB ejecta from dIrrs.
\end{abstract}


\section{Introduction}

The chemical abundance of galaxies  is an integrated information on the sequential process of star formation, stellar death, and ejection of heavy elements lasting over a galactic age under an individual specific galaxy environment. Therefore, an unusual observed chemical feature detected in some galaxies, that is, the feature which is theoretically hard to understand, will tell us some physical key process that drives their chemical evolution and we have overlooked. The low N abundance in H II regions and very young stellar populations of the Magellanic Clouds (MCs) is one of the long-standing puzzling problem that remains unsolved. The relative N/O ratio in the Large MC (LMC) by many studies thus far leads to a mean $\log$ (N/O) $\approx$$-1.4$ -- $-1.5$ \citep[e.g.,][]{Peimbert_74, Dufour_75, Pagel_78, Russell_90, Korn_02, Tsamis_03, vanLoon_10, Rivero_12}, while $\approx$$-1.5$ -- $-1.6$ for the Small MC (SMC) \citep{Russell_90, Pilyugin_03}. Since the solar N/O ratio is $-0.88$ \citep{Asplund_09}, it turns out that the MCs exibit lower N/O ratio by $\sim$$0.5$--$0.7$ dex than the sun.

To assess the origin of low N/O, two observational facts are worth being highlighted. First, the observed N/O ratios in the MCs are broadly equivalent to an average N/O ratio of extremely metal-poor (EMP) stars ([Fe/H]\ltsim$-2.5$) in the Galactic halo \citep{Tsujimoto_11}. Since the chemical composition of EMP stars in the Galactic halo can be explained in the context of SNe-II nucleosynthesis in massive stellar progenitors ($\geq$ 10\msp) \citep[e.g.,][]{Cayrel_04}, it can be interpreted that  N released from asymptotic giant branch (AGB) stars with a time delay, which has lifted the N abundance by $\sim$ 0.6 dex in the solar vicinity,  is missing in the present-day interstellar matter (ISM) of the MCs.

Secondly, the N-deficient feature is in common with many other dwarf irregular galaxies (dIrrs) in the Local Universe, as represented by the well-known plateau of the N/O ratio \citep[$\sim-1.5$, e.g.,][]{Pilyugin_10} for extragalactic H II regions at 12 + $\log$ (O/H) \ltsim 8.0. This fact may suggest that the mechanism to lower the N abundance is associated with the common properties characteristic to these dIrrs. Moreover, neutral gas in bursting dIrrs (i.e., Blue Compact Dwarf (BCD) galaxies) that accounts for a much larger mass in galaxies than ionized one also exhibit low N/O ratios \citep[e.g.,][]{Aloisi_03, Lebouteiller_09, Lebouteiller_13}. This argument holds for the MCs, as implied from low N/O of young MC stars \citep[e.g.,][]{Korn_02, Rolleston_03}.

Theoretically, it is a hard task to predict such a low N/O ratio for dIrrs covering a wide metallicity range up to 12 + $\log$ (O/H) $\sim$ 8.4 for the LMC as not a passing point but a kind of terminus of galaxy evolution.
\citet{Henry_00} show that a low N/O ratio can be identical to the values at the early phase ($\sim$ a first 1-2 Gyr) of predicted evolutionary tracks \citep[see also][]{Vila-Costas_93, Kobulnicky_98, Pilyugin_03, vanZee_06a}. On the other hand, \citet{Molla_06} find that a low N/O ratio can be reproduced as a present-day value as long as 12 + $\log$ (O/H) \ltsim 8.0, utilizing a very low N yield in massive stars. It should be here of note that a low N/O ratio ($\sim -1.5$) extends to 12 + $\log$ (O/H) $\approx$ 8.6-8.7, i.e., a solar metallicity, for dIrrs in the Virgo Cluster \citep{Vilchez_03}. Some models adopting starbursts with burst duration of 0.01 Gyr and/or 0.5 Gyr yield a low N/O ratio \citep[][see also Lanfranchi \& Matteucci 2003; Henry et al.~2006]{Chiappini_03}. In modeling, however, we should be mindful of the star formation history of individual dIrrs. Many studies clearly reveal that the MCs undergo a continuous star formation over more than 10 Gyr \citep[e.g.,][]{Harris_09, Cignoni_12, Weisz_13}. In addition,  since a deep image provided by {\it Hubble Space Telescope} (HST) in particular leads to a color magnitude diagram reaching a very faint magnitude, accurate star formation histories of other dIrrs in Local Volume ($<$ 5 Mpc) are now accessible \citep[e.g.,][]{Tolstoy_09}. Indeed, as discussed in the next section, dIrrs exhibiting low N/O ratios have a long-term star formation similar to the MCs.  

The production site of N in AGB stars is inclined to massive ones such as \gtsim 4 \ms in which a hot bottom burning operates at the bottom of an envelope \citep[e.g.,][]{vandenHoek_97}. Accordingly, AGB stars release N with a time delay of $\sim$0.1 Gyr from their steller birth. This fact implies that the N-rich AGB ejecta are likely to interact with those of type Ia supernovae (SNe Ia). Recent results regarding the delay time distribution (DTD) of SNe Ia yielded by the studies on the SN Ia rate in external galaxies dramatically shorten the SN Ia's delay time, compared with its conventional timescale of $\sim$1 Gyr \citep{Pagel_95, Yoshii_96}. \citet{Mannucci_06} claim that about 50 \% of SNe Ia explode soon after their stellar birth, and further works reveal that the DTD is proportional to $t^{-1}$ with its peak at around 0.1 Gyr extending to $\sim$ 10 Gyr \citep{Totani_08, Maoz_10}. Thus, the current view on SNe Ia is organized such that a majority of SNe Ia explode promptly after the bursting explosions of SNe II (prompt SNe Ia), and the rest gradually emerge with a long interval of Gyrs (slow SNe Ia). Therefore, the timing of N ejection from AGB stars is broadly equivalently to that of the explosion of prompt SNe Ia.

\citet{Tsujimoto_12} discuss the origin of the different [Cr, Mn, Ni/Fe] features between the Milky Way and the LMC, and conclude that their difference can be nicely explained if the ejecta of prompt SNe Ia escape from the gravitational potential of the LMC. They also show that the predicted evolution of [$\alpha$-elements/Fe] under the process of chemical enrichment by massive stars together with slow SNe Ia is in good agreement with the observed trend of the LMC. Their finding suggests that the total number of SNe Ia contributing to the chemical enrichment in the LMC is smaller than that in the Milky Way. This prediction is fully compatible with the implication from the observed high [Ba/Fe] ratios in the LMC as well as in the Fornax dwarf spheroidal (dSph) galaxy, which demands about $\sim$1/3 of the SN Ia rate of the Milky Way \citep{Tsujimoto_11s, Bekki_12}. In addition, recent observed finding of the chemical feature mainly enriched by massive stars, i.e., enhanced [$\alpha$/Fe] ratios, of the neutral medium in BCD galaxies \citep{Lebouteiller_09, Lebouteiller_13} may be suggestive of a preferential removal of SN Ia ejecta. This line of evidence suggests that the energetic ejecta of prompt SNe Ia entrain massive stellar AGB ejecta out of the galaxy potential of dwarf galaxies owing to the coincidental occurrence timing between the two. 

Therefore, the question is raised of what determines the fate of AGB ejecta associated with the bursting prompt SN Ia event. It is no doubt that one primary factor is a galaxy potential. The shallow potential is necessary to strip off AGB ejecta but not a sufficient condition. For instance, IC 1613 and NGC 5152, which are much less luminous galaxies than the SMC, show a signature of AGB enrichment in HII regions as their elevated N/O ratios \citep{Lee_03}. In search for the second key factor, some witness might be obtainable from a well-studied evolution of the MCs.

The MCs have experienced close encounters with the Milky Way with the orbital period of $\sim$ 2.5 Gyr \citep{Bekki_07, Diaz_12}.  Such interactions cause the MCs to lose a part of dark and stellar halo \citep{Bekki_11}. This process is likely to cause the joint stripping of AGB and prompt SNe Ia ejecta. On the other hand, if a dIrr is isolated for its overall  evolution, AGB ejecta seem hard to be stripped since the prompt SNe Ia are unlikely to retain a sufficient energy to push them out of a galaxy potential. This view is supported by the observations that the speed of SN-driven outflows detected in many dwarf galaxies does not exceed the escape velocity of the host galaxy \citep[e.g.,][]{Martin_98, Schwartz_04}.

This paper is organized as follows. We start with a brief review on the properties of dIrrs with stress on interactions, classifying them into two groups in terms of the N/O ratio (\S 2). Then, it is followed by modeling the chemical evolutions of N/O for some dIrrs including the LMC (\S 3). In \S 4, we perform numerical simulations for tracing the fate of AGB ejecta in a dIrr analogous to the LMC under the interactions with the Milky Way to investigate the role of an interaction on the stripping of AGB ejecta from dIrrs.

\section{Observational Witness to the N/O Dichotomy}

\begin{table*}
\begin{center}
\caption{Properties of dwarf irregular galaxies exhibiting low and high N/O ratios}
\begin{tabular}{cccccccc}
\hline\hline
Galaxy & group & D (Mpc) & $M_B$ & 12+log (O/H) & $\Delta$log (O/H) & log (N/O) & comment\\ \hline
\multicolumn{8}{c}{low N/O dIrrs} \\ \hline
Leo A$^1$ & Local Group & 0.79 & -11.70 &  7.38 & $-0.06$ & -1.53$\pm$0.09 & sufficiently small  \\
NGC 6822 & Local Group & 0.50 & -15.20 & \ 8.06$^2$ & +0.09 & -1.60$\pm$0.10 & pass through MW$^3$ \\
IC 4662$^{4}$ & Centaurus & 2.44 & -15.22 & \ 8.09$^{4}$ & +0.12 & -1.50$\pm$0.05 & recent merger$^5$ \\
IZw 18 & beyond LV & \ 19.0$^6$ & \ -15.31$^7$ & \ 7.21$^8$ & $-0.77$ & \ -1.61$\pm$0.06$^8$ & DM stripping \\
NGC 5408$^5$ & M 83 & 4.81 & -16.12 & 7.99 & $-0.11$ & -1.46$\pm$0.05 & outflow$^5$ \\
SMC & Milky Way & 0.06 & -16.31 & 8.13 & 0 & -1.58$\pm$0.15 & inside MW \\
Ho II$^9$ & M 81 & 3.39 & -16.72 & 7.92 & $-0.27$ & -1.52$\pm$0.11 & DM stripping \\
NGC 2366$^{10}$ & M 81 & 3.40 & -17.17 &  7.91 & $-0.35$ & -1.61$^{11}$  \ \ \ \ \ & $\prime\prime$ \\
IC 2574$^9$ & M 81 & 4.02 & -17.46 & 7.93 & $-0.38$ & -1.45$\pm$0.08 & $\prime\prime$ \\
NGC 4395 & CVn & 4.61 & -17.78 & \ \ \ 8.09$^{12}$ & $-0.26$ & -1.52$\pm$0.20 & $\prime\prime$ \\
LMC & Milky Way & 0.05 & -17.91 & 8.37 & 0 & \  \  \ -1.48$\pm0.04$$^{13}$ & inside MW \\ \hline
\multicolumn{8}{c}{high N/O dIrrs} \\ \hline
Pegasus & M 31 & \ \ \ 0.92$^{14}$ & -12.60 & 7.93 &  +0.36 & -1.24$\pm$0.15 & undisturbed outskirts$^{15}$ \\
IC 5152 & field & 2.07 & -14.50 & 7.92 & +0.06 & -1.05$\pm$0.12 & isolated \\
IC 1613 & Local Group & 0.73 & -14.51 & \ \ \ 7.90$^{16}$ & +0.04 & -1.13$\pm$0.18 & no association with MW$^3$ \\
DDO 82$^9$ & M 81 & 4.00 & -14.63 & 7.95 & +0.07 & -1.02$\pm$0.04 &  \\
NGC 5264 & M 83 & 4.53 & -15.90 & 8.66 & +0.59 & -0.57$\pm$0.20 & massive DM \\
NGC 625$^{17}$ & Sculptor & 3.89 & -16.30 & 8.14 & +0.01 & -1.25$\pm$0.03 &  \\
NGC 5253 & Centaurus & 4.00 & -17.38 & \ \ \ 8.30$^{18}$ & 0 & -0.84$\pm$0.10 & only central region$^{19,20}$ \\ \hline
\end{tabular}
\end{center}
{\footnotesize Note. --- $^1$Cole et al. (2007), $^2$Hern\'{a}ndez-Mart\'{i}nez et al. (2009),$^3$Teyssier et al. (2012), $^{4}$Crowther \& Bibby (2009), $^5$van Eymeren et al. (2010), $^6$Marconi et al. (2010), $^7$Gil de Paz et al. (2003), $^8$Lebouteiller et al. (2013), $^9$Croxall et al. (2009), $^{10}$Walter et al. (2008), $^{11}$Gonz\'{a}lez-Delgado et al. (1994), $^{12}$Esteban et al. (2009), $^{13}$Bekki \& Tsujimoto (2010), $^{14}$McConnachie et al. (2005), $^{15}$Kniazev et al. (2009), $^{16}$Bresolin et al. (2007), $^{17}$Cannon et al. (2005), $^{18}$L\'{o}pez-S\'{a}nchez et al. (2007), $^{19}$Monreal-Ibero et al. (2012), $^{20}$Westmoquette et al. (2013).}
\end{table*}

In this section, we categorize dIrrs in Local Volume into the ones in which massive AGB contribution is basically unseen for the N abundance in the ISM and others exhibiting elevated N abundances. As the boundary of this selection, we set $\log$ (N/O) $\leq-1.45$ and $\geq-1.25$, respectively, considering a margin of 0.2 dex between the two associated with individual observed errors in N/O ratio. The sample  of dIrrs with their properties is selected from the list compiled by \citet{vanZee_06}. We mainly focus on the data which have the O abundance directly measured using [O III] $\lambda$4363 line. Exceptionally, a well-studied IZw 18 beyond Local Volume is added. The updates for some data are adopted with their source references indicated. The final list is shown in Table 1. The sixth column, $\Delta$log(O/H), means a deviation from the observed metallicity-luminosity relation within the Local Volume. Here we adopt the relation 12+$\log$ (O/H) = 5.67 - 0.151$M_B$ \citep{vanZee_06}. Since this relationship can be interpreted as a result of the control of an enrichment achievement by the gravitational potential wells \citep[e.g.,][]{Tremonti_04},  we regard a large deviation as an evidence for a peculiar dark halo such as a massive dark halo (+$\Delta$) or the striping of an outer part of dark halo through a tidal interaction($-\Delta$). On the other hand, a small deviation yielding a scatter of this relation is possibly regarded as a result of phenomena specific to individual galaxies. For instance, bursts of star formation that have occurred in the recent past may lead to a small +$\Delta$ due to a prompt O production. Here it should be of note that the direct method of O abundance analysis does not support a level of accuracy for O/H better than 0.2 dex \citep[e.g.,][]{James_13a}. In light of this fact, the values of $\Delta$ less than 0.2 may not be taken at face value.   

\begin{figure}
\includegraphics[width=230pt]{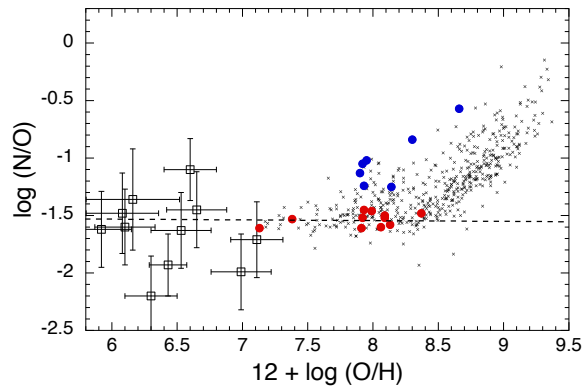}
\caption{Correlation of N/O with O/H for dIrrs listed in Table 1 (red circles: low N/O dIrrs, blue circles: high N/O dIrrs) together with those for Galactic halo stars \citep[squares:][]{Spite_05}. For reference,  the data of extragalactic H II regions including those of bright massive galaxies assembled by \citet{Pettini_08} are shown by small crosses. The dashed line is the average ratio of low N/O dIrrs denoted by red circles.}
\end{figure}

Figure 1 shows the N/O ratios selected for two groups in Table 1, i.e., low N/O dIrrs (red circles) and high N/O dIrrs (blue circles), compared with those of extremely metal-poor stars in the Galactic halo (squares). We see that the level of N/O for low N/O dIrrs is broadly compatible with that of Galactic halo stars, which is considered to reflect the nucleosynthesis N/O ratio in massive stars. In addition, for reference, much more data of extragalactic HII regions compiled by \citet{Pettini_08} are denoted by small crosses. These data include N/O for bright massive galaxies such as M 31, M 83, M 51, and NGC 4254 which mainly populate at 12 + $\log$ (O/H)$>$9.0.

\subsection{Low N/O dwarf irregular galaxies}

Leo A is one of the lowest luminous gas-rich galaxies. Since a deep image provided by HST observation reveals its star formation history, which has continuously lasted at least 8 Gyr \citep{Cole_07}, Leo A is a good example of a dIrr exhibiting a low N/O ratio regardless of a long-term star formation. Its low N/O ratio is likely attributed to the galaxy mass which is so small that massive AGB ejecta can easily escape from a galaxy potential. NGC 6822 is also found by HST color-magnitude studies to have a long star formation history over $>$10 Gyr \citep[e.g.,][]{Wyder_03, Cannon_12}. Though NGC 6822 is currently isolated from either the Milky Way or M 31, its observed distance and velocity in the context of a cosmological simulation suggests that this galaxy belongs to the population which has passed the Milky Way, together with Sextans A and Sextans B exhibiting $\log$ (N/O) = $-1.54\pm0.13$, $-1.46\pm0.06$, respectively \citep{Teyssier_12}. On the other hand, gas dynamics of two Magellanic-type galaxies, IC 4662 and NGC 5408, is well investigated, and expanding superbubbles are detected for both galaxies as well as the feature of twisted velocity field in IC 4662, likely caused by a recent merger event \citep{vanEymeren_10}. The M 81 group is dynamically unstable, leading to tidal interactions among galaxies that allow the presence of tidal dwarf galaxies \citep[e.g.,][]{Makarova_02} and tidal streams and debris \citep[e.g.,][]{Yun_94}. Its member galaxies, Ho II, NGC 2366, IC 2574, exhibit significantly lower O abundances by $\sim$ 0.2-0.4 dex than those expected for their corresponding luminosities. That might be a result of stripping of dark halo through tidal interactions. Note that all of three galaxies have long-term and broadly constant star formation histories \citep{Weisz_08}. Likely, a low-luminosity Seyfert galaxy NGC 4395 in the Canes Venatici Cloud (CVn) group may have a similar stripping history, though its O abundance involves a large uncertainty including the LMC-like high value \citep{vanZee_98}. In addition to them, IZw 18 is likely to be significantly stripped off a dark halo, which may result in the record as the most metal-deficient galaxy known. This star-forming galaxy is found by HST photometry to be composed of stars of all ages possibly including 13 Gyr-old stars \citep{Ramos_11}. Another sample of dIrr beyond the Local Volume, UM 462, can be added to this category since this galaxy ($D$=14.4 Mpc) is a disrupted system showing sings of interaction in the emission-line map and exhibits $\log$ (N/O)=$-1.85\pm$0.15 with 12 +$\log$ (O/H)=8.03$\pm$0.10 \citep{James_10}.

If we can measure the N/O ratio of a blob of stripped N-rich ejecta in the interacting dIrrs, its value must be high, and becomes more direct evidence for stripping of N-rich ejecta from low N/O dIrrs. Here are two dIrrs potentially corresponding to this case. UM 448 ($D$=75.9 Mpc) is an interacting system of 2 younger/older bodies with evidence of elevated N/O ratio ($\log$ N/O = $-1.03\pm0.09$) at the zone of interaction \citep{James_13b}. Haro 11($D$=83.6 Mpc) is an interacting Lyman-break analogue of 3 bodies, with one knot showing lower O and higher N/O ($\log$ N/O = $-1.12\pm0.05$) than the rest (e.g., $\sim-1.41$) of the system \citep{James_13a}.

\subsection{High N/O dwarf irregular galaxies}

Pegasus has an extended ($\sim$ 8 kpc) stellar disk that has no expected distortions caused by a tidal interaction \citep{Kniazev_09}. The luminosity of this galaxy might be sufficiently low to have low N/O as in the case of Leo A. However, Pegasus seems to possess a massive dark halo, which is inferred from a higher O abundance than predicted for its luminosity \citep[see also][]{Skillman_97}. In the end, an isolated environment with a massive halo is likely to result in a relatively high N/O. Possession of a massive dark halo is also implied for NGC 5264 from a large $\Delta\log$ (O/H) of +0.59. IC 1613 is a member of Local Group, but the likelihood of association with the Milky Way (passing through the virial radius) is estimated to be less than 1 percent \citep{Teyssier_12}. NGC 5253 is well studied about the distribution of O and N abundances inside the galaxy. It has been found by several authors \citep[e.g.,][]{Monreal_12, Westmoquette_13} that both O and N abundances are nearly constant except for the central \ltsim 50 pc region where the N/O ratio is significantly elevated owing to high N abundances. Since the constant N abundance yields $\log$ (N/O) $\sim$ -1.5 \citep{Lopez_07, Westmoquette_13}, NGC 5253 might be suitable for the category of low N/O dIrrs. A more extreme case than NGC 5253 with N-enriched dense core can be seen in Mrk 996, which is an isolated dIrr at a distance of 22.3 Mpc \citep{James_09}. This galaxy exhibits $\log$ (N/O)=$-0.13$ in the core within a diffuse N-poor envelope ($\log$ N/O=$-1.43$). 

\section{Chemical Evolution with AGB stripping}

\begin{figure*}
\includegraphics[width=320pt]{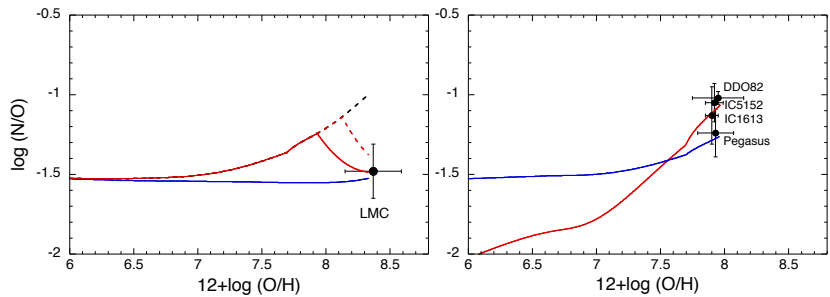}
\caption{{\it Left panel}: N/O evolution predicted by the LMC model, varying an assumption regarding AGB stripping, i.e., AGB stripping occurs over the whole evolution (blue line), except for the first 5 Gyr (red solid line) or 8 Gyr (red dashed line), and does not occur (black dashed line). {\it Right panel}: N/O evolution predicted for dIrrs exhibiting high N/O at a low metallicity. AGB contribution to N enrichment is fully included in the models with a steep IMF (blue line) and a truncated IMF at 20 \ms (red line).}
\end{figure*}

Incorporating the possible stripping of AGB ejecta into the chemical evolution models, we examine the evolutionary paths of N/O for dIrrs. Since we already construct a successful model for the evolutions of C/O and N/O in extragalactic H II regions \citep{Tsujimoto_11}, we utilize the same model formula as in the previous paper except for SN Ia model, together with a new treatment of AGB ejecta. We adopt the new formula for SNe Ia that are composed of prompt and slow SN Ia \citep{Tsujimoto_12}, though this study is unaffected by SN Ia formula owing to a negligible (no) contribution of O (N) from SNe Ia to chemical enrichment. Here we refer to the adopted yield of N. For the nucleosynthesis N yields of massive stars, we use the empirical yield deduced by \citet{Tsujimoto_11}. The derived yield is based on the observed N/O ratios of extremely metal-poor stars \citep{Spite_05} as well as recent nucleosynthesis results for rapidly rotating massive stars giving high N production \citep{Hirschi_07, Ekstrom_08}. For the AGB yield, we assume a metallicity-dependent N yield for stars with the masses of 1 - 8 \msp. As a basic yield, we adopt the results calculated with $Z$=0.001, $n_{\rm AGB}$=4, and $m_{\rm HBB}$=0.8 by \citet{vandenHoek_97}, and we choose its metallicity dependence so as to follow the observed increasing feature in N/O for extragalactic H II regions as shown in Figure 1. 

Slow star formation proceeded in dIrrs is parameterized by a low SFR coefficient, $\nu$ = 0.08-0.25 Gyr$^{-1}$ (the mass fraction of the gas converted into stars per Gyr, and note that $\nu$ = 0.4 for the solar neighborhood), adjusting the final O abundance to each observed one in H II regions. For all galaxies, the galactic age of 13.5 Gyr is assumed.  

\subsection{LMC}

Among low N/O dIrrs, we calculate a few possible paths of N/O evolution for the LMC including the effects of AGB stripping. For the initial mass function (IMF), we adopt a steep IMF ($x$=$-1.6$), which is predicted by previous studies on the chemical evolution of the LMC \citep{Tsujimoto_12}. First, we take a look at the model result without AGB stripping over the whole evolution (black dashed line in the left panel of Fig.~2). Owing to a gradual  increase in N abundance caused by continuous enrichment by AGB  stars, the predicted present-day N/O is largely deviated form the observed value. On the other hand, an assumption that massive (4 - 8 \msp) AGB stars do not contribute to all chemical enrichment history gives a good agreement with the observation to the result (blue line). More realistic cases for AGB stripping are further considered. Numerical simulations of AGB stripping analogous to the LMC case presented in the next section suggest that during the first $\sim$5 - 8 Gyr, AGB ejecta is kept within an inner halo of the LMC.  Taking this numerical result into consideration, we exemplify two simple cases  that the ejecta of massive AGB  stars can be recycled for the first 5 (red solid line) or 8 Gyr (red dashed line) while no contribution to chemical enrichment is assumed for the latter evolution. Both results are compatible with the observed N/O ratio within an error bar. 

These results suggest that the present-day low N/O can be explained as long as stripping of massive AGB ejecta has  occurred at a very high rate during at least last several Gyrs.  Moderate stripping would yield a higher N/O than that expected from the observation. According to our current numerical simulations, a sufficient stripping efficiency while the stripping process is activated is realized only for the model with less massive dark halo (see \S 4). Since such high AGB stripping is essential to achieve the low N/O as in the LMC, its process in the framework of dynamical evolution should be more precisely and in further detail validated by our future simulations with high resolution.  

\subsection{High N/O dIrrs}

Here we treat the most interesting case where a high N/O ratio is realized at a very low metallicity. Four dIrrs, i.e., Pegasus, IC 5152, IC 1613, DDO 82, exhibit $\log$ (N/O) = $-1.24$ to  $-1.02$ at a similar metallicity of 12+$\log$(O/H)$\approx$ 7.9. These high N/O cases are examined with an assumption that AGB contribution is fully included together with a steep IMF ($x$=$-1.6$) as in the same for the LMC. The model result is shown by blue line in the right panel. We see that it predicts a relatively low level of N/O as compared with the observed values among four dIrrs. The model calculated with a Salpteter IMF ($x$=$-1.35$) lowers a reaching N/O by $\sim$ 0.1 dex and thus leads to deviation from the observed point. More enhanced N/O ratio can be obtained by adopting a different form of the IMF, that is, the IMF with a cutoff at the upper mass end, $m_u$,  at around 20-25 \msp, which is implied by high-z metal-poor damped Ly$\alpha$ (DLA) systems \citep{Tsujimoto_11} as well as by some dSph galaxies such as the Fornax \citep{Tsujimoto_11s}. This is because such a IMF lacks very massive stars which synthesize O in large quantities, while the contribution of N from AGB stars remains at the same level (note that a steep IMF reduces both contributions). As expected, the result calculated with $m_u$=20 \ms and $x$=$-1.35$ is found to be in better agreement with the observed values by reaching a higher N/O (red line). This predicted path for 12+$\log$(O/H) \ltsim 7.5 coincides with the N/O ratios of high-z  DLAs \citep{Pettini_08}. We, however, do not exclude the possibility that such high N/O ratios are affected to some extent by local contamination from the ejecta of Wolf-Rayet stars \citep[e.g.,][]{Lopez_10}, 
in particular for the extremely high N/O dIrrs such as Mrk 996 \citep{James_09}.

\subsection{Broad view on N/O among dIrrs}

Except for the unusually high N/O ratios as exhibited by the three galaxies (IC 5152, IC 1613, DDO 82), the observed N/O ratios of dIrrs for 12+$\log$ (O/H) \ltsim 8.5 are basically resided within the triangular-shaped cloud of data points, the vertices of which are (12+$\log$ O/H, $\log$ N/O)= (7.21, $-1.61$), (7.93, $-1.24$), (8.37, $-1.48$), corresponding to IZw 18, Pegasus, and the LMC, respectively (see Fig.~1). Since the highest N/O ratio among them is broadly identical to the one of Pegasus, which is modeled with an assumption of a complete contribution of AGB ejecta to chemical enrichment, a rather small scatter of the N/O ratio in the low-metallicity regime can be explained by the process of AGB stripping that can happen at different times and with different intensities. It should be here  stressed that the model results in Figure 2 can only be compared to the present-day (i.e., end-point) HII region abundances of many other dIrrs, while a comparison with N/O ratios prior to this epoch is impossible due to the lack of suitable diagnostics, such as abundances in dim long-lived stars. Accordingly, the overall interpretation of the N/O ratios exhibited by dIrrs can be summarized as follows; (i) a plateau-like low N/O of $\sim -1.5$ is attributed to an efficient stripping of AGB ejecta due to a sufficiently small galaxy mass or a strong interaction with other massive galaxy as experienced by the LMC, and (ii) a confinement of AGB ejecta within a galaxy potential or their moderate stripping eventually lifts a N/O ratio to some extent, $\sim$ 0.3 dex at most. 

A common property of dIrrs including unusually high N/O ones in our scenario is their IMFs which favor a small number of massive stars. This is an outcome of chemical evolution that demands less enrichment by massive stars to the ISM in dIrrs, and may be suggestive of the fact that partial removal of SN II ejecta is ubiquitous in dIrrs, as will be discussed in \$ 5.1.

\section{Numerical simulations of tidal stripping of AGB ejecta}

We here investigate how much of gas released from AGB stars can be stripped from dwarf galaxies orbiting a luminous disk galaxy by using N-body + hydrodynamical simulations. In order to perform numerical simulations on GPU clusters, we have revised our previous simulation code \citep[GRAPE-SPH;][]{Bekki_09}. The present model is  a hybrid one in the sense that the two-hold dynamical model (i.e., orbital model plus N-body model) constructed to simulate an interactive evolution between the Milky Way and the MCs \citep{Diaz_12} is combined with a hydrodynamical code including star formation and supernova feedback effects \citep{Bekki_13}. Details are described in the two papers. We briefly describe the model applying to the present study.

\subsection{Orbital model}
We determine the orbital evolution of a dwarf galaxy with respect to the Milky Way for the last $\sim$ 11 Gyr, using a backward integration scheme. The Milky Way influences the orbits of a dwarf galaxy through a fixed gravitational potential of three components: a central bulge, a disk, and an extended dark matter halo.  Each component is built to give the best model for the Magellanic Stream formation \citep{Diaz_12}. A dwarf galaxy is assumed to have a Plummer potential with a different mass of $M_{\rm dw}=3.0 \times 10^{10} M_{\odot}$, $4.5 \times 10^{10} M_{\odot}$, and $6.0 \times 10^{10} M_{\odot}$. In addition, dynamical friction is assumed to operate separately on the orbit of a dwarf galaxy as it passes through the Milky Way's dark halo, following the Chandrasekhar formula \citep{Binney_08}.

We carry out our simulations in a galactocentric frame in such that the center of the Milky Way is fixed at ($X$,$Y$,$Z$) = (0,0,0).  The assumed initial position vectors and space velocities of a dwarf galaxy are the same as those adopted for the LMC, i.e., ($X$,$Y$,$Z$) =  (-0.8, -41.6, -27.0) kpc and ($U$,$V$,$W$) = (-50.7, -226.1, 229.3) km s$^{-1}$ \citep[see Table 1 in][]{Diaz_12}. In our model, the orbital evolution of a dwarf galaxy depends solely on $M_{\rm dw}$. Figure 3 shows the orbital evolutions with three different $M_{\rm dw}$. 

\begin{figure}
\includegraphics[width=230pt]{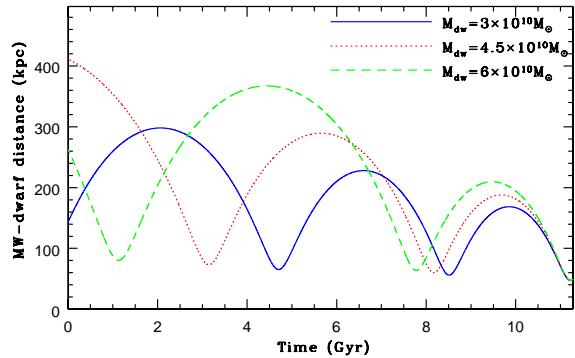}
\caption{Time evolution of the distance between a dwarf galaxy and the Milky Way in the last 11 Gyr for three different dwarf models with $M_{\rm dw}=3 \times 10^{10} M_{\odot}$ (blue solid line), $M_{\rm dw}=4.5 \times 10^{10} M_{\odot}$ (red dotted line), and  $M_{\rm dw}=6.0 \times 10^{10} M_{\odot}$ (green dashed line).
}
\end{figure}

\subsection{N-body+hydrodynamical simulations}

For the determined orbits of dwarf galaxies, we trace the motion of AGB ejecta within a modeled dwarf galaxy where star formation proceeds. A dwarf galaxy is represented as a bulgeless disk galaxy embedded in a massive dark matter halo. The mass- and size-ratios of dark matter halo to initial stellar disk are set to be 16.7 and 14, respectively. The scale length and size $R_{\rm d, dw}$ of stellar disk are assumed to be 1.5 kpc and 7.5 kpc, which are analogous to the disk structure of the LMC \citep{Bothun_88}. Thus, the dark halo has a size of 105 kpc. 

The gas disk is assumed to be exponential and have a mass of $2.3M_{\rm d, dw}$ with its size of $2R_{\rm d, dw}$. The total energy of individual SN II and SN Ia  is set to be $10^{51}$ erg, and its 90\% is used for thermal feedback while the remaining is used for kinematic feedback. Each form of the deposited energy increases the thermal energy (i.e., temperature) and random motions of the surrounding ISM, respectively. The total numbers of particles used for dark matter, stellar disk, and gaseous one are 400,000, 100,000, and 100,000, respectively. In the following,  $T$ represents the elapsed time since a start of simulation.

\subsection{Results}

First, we see the detailed results calculated by the model with $M_{\rm dw}=4.5 \times 10^{10} M_{\odot}$. The upper panels of Figure 4 demonstrate that gas initially in the disk of a dwarf galaxy is stripped and then
distributed within the outer halo of the Milky Way through an orbital evolution. Owing to SN feedback effects, some fraction of gas in the disk is ejected into the dwarf's halo, and becomes susceptible to tidal stripping. Then real stripping occurs especially when a dwarf galaxy passes its orbital pericenter at $T \sim 3$ and 8 Gyr, since a large fraction of its dark matter halo is stripped by the strong tidal field of the Milky Way. At the first pericenter passage, a small fraction of gas ejected into the dwarf's halo is stripped to form diffuse gas streams, which will not return to the ISM in the dwarf's disk. After the second pericenter passage, a dwarf galaxy loses a much larger fraction of gas under the truncated dark halo. The stripped gas finally shows a thin ring-like feature. Interestingly,  some fraction of its gas is accreted onto the inner region of the Milky Way.

\begin{figure}
\includegraphics[width=230pt]{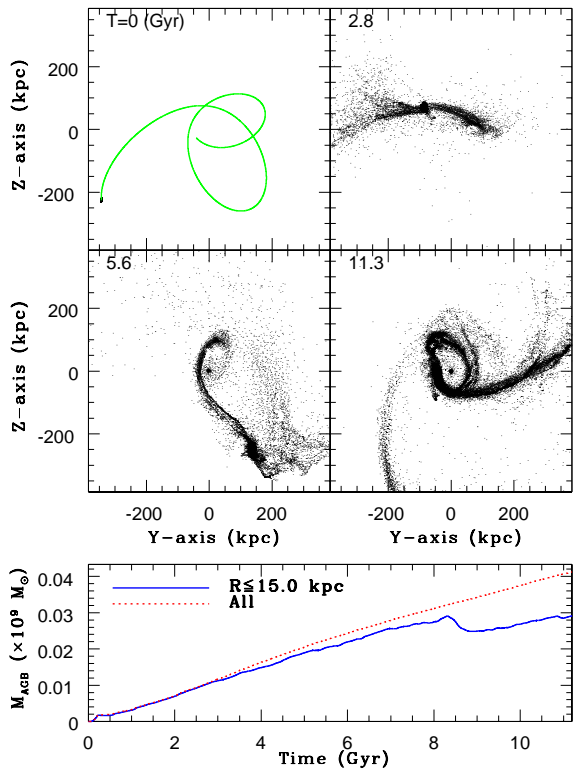}
\caption{{\it Upper panels}: The mass distribution of gas projected onto the $y$-$z$ plane at different four time steps for the model with $M_{\rm dw}=4.5 \times 10^{10} M_{\odot}$. The time $T$ shown in the upper left corner for each panel is given in units of Gyr. The orbit of a dwarf galaxy is indicated  by thick green line in the upper left panel for $T=0$ Gyr. {\it Lower panel}: Time evolution of the mass of AGB ejecta for the same model as in upper panels. The mass of AGB ejecta ($M_{\rm AGB}$) within $R \le 15$ kpc and that of all AGB ejecta are shown by red dotted and blue solid lines, respectively.
}
\end{figure}

The lower panel of Figure 4 shows the mass $M_{\rm AGB}$ of AGB ejecta within 15 kpc of a dwarf galaxy halo, compared with the total mass of gas ejected from AGB stars. The difference between the two is seen after the first pericenter distance at $T \sim 3$ Gyr, and $M_{\rm AGB}$ within a halo clearly drops owing to a tidal stripping of AGB ejecta from the halo after the second pericenter passage at $T \sim 8$ Gyr. In this model, 28\% of all gas ejected from AGB stars is finally stripped from the dwarf's halo with no return back to the original gas disk up to the present-day. Provided that the stripped ejecta originate in the upper mass range of AGB progenitors as implied from the chemical studies, this fraction will meet the condition that N released from massive AGB stars is significantly stripped off and thus do not contribute to an enrichment in the ISM. Since we aim at tracing an overall evolution of a dwarf galaxy under an interactive environment, we are forced to discard accurate resolution to simulate the behavior of individual ejecta released from SNe II, prompt and slow SNe Ia, and low-mass and massive AGB stars. Accordingly, our result is limited to show that the ensemble of AGB ejecta are lifted up to the halo by feedbacks from the assembly of numerous SNe II and Ia (see also discussion in \S 5.1). 

\begin{figure}
\includegraphics[width=230pt]{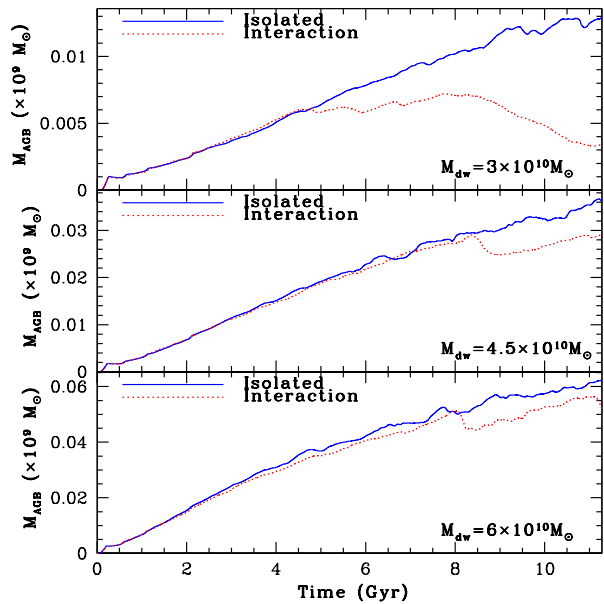}
\caption{Time evolution of the total mass of AGB ejecta  ($M_{\rm AGB}$) within 15 kpc of a dwarf galaxy halo for  the isolated (blue solid line) and interaction (red dotted line) models for three different dwarf masses; $M_{\rm dw}=3 \times 10^{10} M_{\odot}$ (top), $M_{\rm dw}=4.5 \times 10^{10} M_{\odot}$ (middle), and  $M_{\rm dw}=6.0 \times 10^{10} M_{\odot}$ (bottom).
}
\end{figure}

Next we present the importance of tidal interaction to strip AGB ejecta, introducing the isolated model where the gravitational influence of the Milky Way on dwarf galaxies is not included. Figure 5 shows the comparison of time evolution of $M_{\rm AGB}$ within a halo for three different $M_{\rm dw}$ between the models calculated under an orbital evolution presented in Figure 1 and the isolated ones. For all of three models, differences in the evolution of $M_{\rm AGB}$ within a halo are seen, in particular, a factor of $\sim 4$ difference in the final mass  for  the lowest mass model. On the other hand, the highest mass model retains most of AGB ejecta in a dwarf halo even after close interactions at  $T \sim$ 1 and 9 Gyr. These results are attributed to a different extent of dark halo stripping in the sense that a less massive dwarf galaxy loses more dark halo through tidal interactions. 
The derived dependence of the removal efficiency on the mass strongly suggests that N stripping should be efficient in low-mass, low-metallicity galaxies and much less efficient in high-mass, high-metallicity galaxies, which is compatible with the observed N/O tendency as shown in Figure 1. In the end, we claim that the chemical evolution of dwarf galaxies is largely influenced by their galaxy surroundings, i,e, satellite galaxies in a luminous galaxy or an isolated environment, together with by their dynamical masses. 

\section{Discussion}

\subsection{Removal efficiencies of ejecta between AGB, SN Ia, and SN II}

Our theoretical scheme is constructed under a hypothesis that preferential loss of massive AGB ejecta together with those of prompt SNe Ia occurs in dIrrs. This view is favoured by hydrodynamical simulations in the starburst model \citep{Recchi_01}. In this model, SNe II explode in dense ISM where most of energy from SNe II is dissipated away by radiative cooling owing to the high-density environment \citep[e.g., 97\%;][]{Bradamante_98}, and the rest of energy is used for the formation of expanding shells, which is responsible for the formation of HI hole or very diffuse ISM. Afterward, SNe Ia explode and AGB ejecta are released in an environment with little ISM where these ejecta can be expelled from the disk driven by SNe Ia's explosion energy. We suppose that an efficient joint mass loss from prompt SNe Ia and massive AGB stars is realized even under a continuous star formation by regarding it as a sequence of small-scale starbursts. In other words, global continuous star formation is an ensemble of localized small bursts. In each local region, stars are intermittently formed at intervals of a few $10^8$ yrs. In localized small bursts of star formation, the ambient ISM is first thermalized by SN II explosions so as to suppress further star formation within $\sim10^8$yrs, then followed by feedback of prompt SNe Ia, and finally the ISM can cool down without a major heat source. Such an interval of star formation is supported by the discussion on the turbulent ISM following SN explosions. \citet{Avillez_02} discuss the mixing timescales of SN II ejecta and quote a shortest mixing timescale of $1.2\times10^8$yr for the ambient ISM. It might predict an unacceptably large scatter in stellar abundances due to the incomplete chemical homogeneity of the ISM unless we adopt the interval of star formation more than a few $10^8$yr.

This scheme will be validated by performing highly-resolved hydrodynamical simulations applied to local regions where individual stars and ejecta can be decomposed, extracted from an entire galaxy (Bekki \& Tsujimoto, in preparation). We believe that such future numerical simulations can shed light on the connection between the star formation and the removal efficiencies among the different heavy-element ejectors. Observationally, there is evidence for bursting star formation even in the Milky Way's past over most of cosmic time. \citet{Rocha-Pinto_00} find that the local disk within $\sim 100$ pc has experienced a complex star formation history, using a chromospheric age distribution of dwarf stars. We will see that the LMC has a much more complex one with the strong SFR bumpiness on account of its small mass if we can have a close look at stellar properties of the individual LMC stars, and thereby deduce the localized, not averaged, star formation history. 

We do not object to removal of SN II ejecta from a galaxy potential of dIrrs. It is reasonable to anticipate that the ejecta of SNe II is in part removed from dwarf galaxies. As  supporting evidence, galactic winds from dwarf starburst galaxy, NGC 1569, exhibit an enhancement of  $\alpha$-elements relative to Fe which is indicative of a SN II-like feature \citep{Martin_02}. Note that this galaxy shows a marginally low (N/O) ratio of $-1.39\pm0.05$ with 12+$\log$ (O/H) = 8.19 \citep{Kobulnicky_97}. Additionally, two dwarf starburst galaxies (NGC 4449, He 2-10) are likely to retain the hot gas with an enhanced $\alpha$/Fe while a non-enhanced feature is seen for two galaxies (NGC 3077, NGC 4214) \citep{Ott_05}. As already discussed, chemical evolution models for dIrrs including the LMC support a steeper IMF. This required reduction of SN II enrichment can be  identical to the outcome of a partial removal triggered by galactic winds associated with SNe II. This angle is supported by the observational studies on the IMF of the MCs claiming a Salpeter-like IMF \citep[e.g.,][]{Bastian_10}. An IMF slope of $x$=$-1.6$ adopted in the present study indeed broadly corresponds to a 50\% removal of SN II ejecta. It turns out that the differences in removal efficiencies among SNII, SN Ia, and AGB ejecta may be quite moderate. Recall here that we assume a removal of subset (not all) of  both AGB (only massive ones) and SN Ia (only prompt ones) ejecta. These similar efficiencies are rather favored by previous works \citep{Recchi_01, Recchi_02, Recchi_04}. In any event, the degree of escape of SN II ejecta is of secondary importance for low N/O dIrrs focused in this study: the low N/O ratios for dIrrs such as the LMC mimic the effects of SN II nucleosynthesis, though the underlying cause is the preferential stripping of N-rich AGB ejecta. On the other hand, the N/O evolution of dIrrs that can confine AGB ejecta is largely influenced by a degree of SN II removal since the N/O track goes up by the combined effects of N enrichment by AGB stars and less O contribution by SNe II.

The loss of AGB ejecta from dIrrs is restricted to those originated from massive (\gtsim 4 \msp) AGB stars since prompt SNe Ia play a key role of their thermalization and eventual removal. Therefore, in our scenario, $s$-process elements such as Ba synthesized in AGB stars with a mass of 1.5 - 3\ms \citep{Busso_01} are hard to suffer a removal from dIrrs. On the contrary, the observed results present that stars belonging to the LMC as well as to some dSph galaxies exhibit unusual high Ba abundances \citep{Pompeia_08, Sbordone_07, Letarte_10}.  On the other hand, since the production site of C is inclined to a more massive star, i.e., a 2 -4 \ms AGB star \citep{vandenHoek_97}, a partial removal of C ejected form AGB stars is expected. Indeed, the MCs show the deficiency of C in H II regions but to a lesser extent than N \citep[e.g.,][]{Hill_04}.

\begin{figure}
\includegraphics[width=230pt]{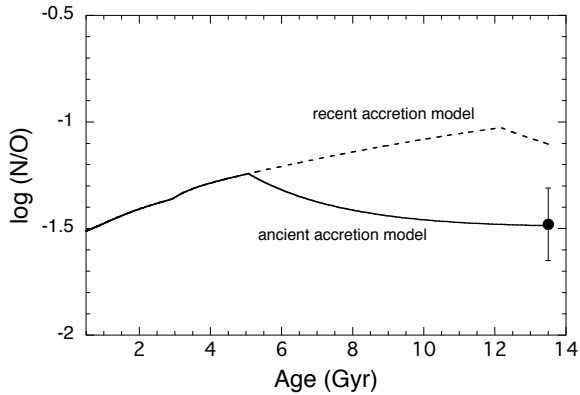}
\caption{N/O evolution as a function of a galactic age predicted for the LMC. Two models are assumed to have a phase where massive AGB stars do not enrich the ISM by N due to the stripping of AGB ejecta caused by the interaction with the Milky Way during the last 1.5 Gyr (recent accretion model) and 8.5 Gyr (ancient accretion model), respectively.
}
\end{figure}

\subsection{Pinpointing an accretion epoch of the LMC}

The N/O ratio as diagnosis of a galaxy environment as we propose can be applied to discussion on the debatable motion of the LMC. Recent measurement of the proper motion of the LMC using HST suggests that the LMC is now in the middle of a first infall to the Milky Way \citep{Kallivayalil_06, Kallivayalil_13}. Their canonical orbit implies that the LMC first entered within 100 kpc of the Milky Way only a few $10^8$ yr ago. Here we calculate the N/O evolution in the LMC for the case analogous to this first infall model. We assume that N enrichment by massive AGB stars ceases from 1.5 Gyr ago corresponding the accretion epoch reaching a virial radius (300 kpc) of the Milky Way. As shown in Figure 6, the recent accretion model does not give the sufficient time to lower the N/O ratio down to the present value (dashed line), while the periodic orbit model which has the  orbital period of $\sim 4-5$ Gyr within a virial radius matches the observation (solid line), as already discussed in \$ 4. These results suggest that the LMC has approached the Milky  Way at least more than a few Gyr ago.

This approach can be strengthened by acquiring stellar N/O for unevolved stars in the LMC. The N/O ratios for stars with different metallicities will give information on how the N/O ratio changes with metallicity, which acts as  an indicator of age. This N/O-age relation thus obtained will enable us to pinpoint an accretion epoch of the LMC, though it is currently a formidable task to measure the N abundance for dim unevolved stars in Local Group. In future, the same method will be also applied to infer the epoch of accretions of dSph galaxies onto the Milky Way.

\subsection{N/O in different galaxy environment}

In cluster of galaxies, dwarf galaxies may be inclined to exhibit low N/O owing to an efficient stripping of AGB ejecta by strong tide as well as ram pressure of hot intra-cluster medium. In the Virgo Cluster, $\sim$70\% of observed 21 dIrrs exhibit low N/O ratios \citep{Vilchez_03}. This percentage derived from a small database, in fact, seems not so high since 50 dIrrs in the Local Volume listed by \citet{vanZee_06} involves $\sim$ 80\% low N/O dIrrs (however, $\sim55\%$ shows $\log$ (N/O)$<-1.45$ for 12+$\log$ (O/H)$<$8.2 in extragalactic H II region data  compiled by Pettini et al.~2008). In addition, among the member dIrrs in the Virgo, we do not see a clear correlation between a low-N/O cluster member and its location relative to the centre of the cluster (i.e., M 87).  Given the absence of a clear correlation between them, a scenario involving ram stripping of cold galactic gas may lead to an increased N/O ratio \citep[e.g.,][]{Marcolini_03, Marcolini_04}, contrary to our main thesis.  This ram-stripping process results in a suppression of further star formation and production of O, followed by a delayed ejection of N. It is indeed different from our proposed stripping of halo gas, where star formation should be negligible. More data for the reliable statistical analysis will be surely demanded to validate how the N/O ratio is associated with a galaxy environment. In Local Group, we predict that N/O ratios in dSph galaxies are essentially low as in the MCs because of their much closer distance to a luminous galaxy, i.e., the Milky Way or M 31 than dIrrs. 

\subsection{Crosscheck of other scenario}

In our proposed stripping scenario, severe truncation of recycling of AGB ejecta due to tidal stripping of gas
from dIrrs is responsible for the origin of low N/O. On the other hand, \citet{Bekki_10} propose an alternative infall scenario in which the observed unusually low N/O in the young populations of the LMC is due to the infall of gas with very low N/O from the SMC in the predicted scheme of a gas-transfer from the SMC to the LMC \citep[e.g.,][]{Diaz_12}. It is possible to consider that similar dilution occurred in some dIrrs by an infall of the objects analogous to H I high velocity clouds (HVCs) that surround the Milky Way. Galactic HVCs are observed to have low [N/H] ranging from $-2.0$ - $-1.2$, giving $\log$ (N/O)$\sim -1.6$ \citep{Collins_07}.This infall scenario predicts a low N/O ratio only for young stars and H II regions. Therefore, a way to distinguish between the two scenarios is to measure the N/O ratio for intermediate-age stars in dIrrs.

\section{Conclusions}

We have assessed the origin of low N/O ratios quite commonly seen for H II regions in dIrrs including the MCs, which can be interpreted as lack of enrichment of N by massive AGB stars. At various level, we find a signature indicating past tidal interactions for the properties of these dIrrs, along with a strong argument that the MCs have closely interacted with the Milky Way. These findings lead to our claim that the ejecta of massive AGB stars are efficiently stripped off from a shallow gravitational potential of dIrrs during their dynamical interaction with their larger host galaxies. Theoretical basis of this mechanism is that massive AGB stars release their ejecta after $\sim$0.1 Gyr from a stellar birth, which broadly coincides with the delay time of the bursting explosion of prompt SNe Ia. These ejecta thus dispersed into the halo are eventually stripped off if dIrrs suffer tidal interactions. Our numerical simulations on the dynamical evolution of dIrrs around the Milky Way have confirmed that the two-stage process (ejection in halo followed by removal) for efficient stripping of AGB ejecta really occurs in dIrrs. 

Our new chemical evolution models predict various paths of N/O evolution for dwarf galaxies according to the possible diversity of AGB stripping history that results form past tidal interactions. Future measurements of N/O abundance ratios for unevolved turn-off stars as a function of metallicity for nearby dwarf galaxies by the upcoming next-generation telescopes will open a new paradigm where the chemical evolution is discussed in close connection with the effects of galaxy environments. 

\section*{Acknowledgments}
T.T. is grateful for ICRAR's hospitality during a research visit at which this work was completed. The authors wish to thank the referee for all valuable comments that helped improve the paper. Numerical computations were carried out at the University of Western Australia on a GPU cluster implementing the CUDA G5/G6 software package for calculations of gravitational dynamics. This research was supported by the Graduate University for Advanced Studies (SOKENDAI).

\label{lastpage}

\end{document}